\documentstyle[12pt]{article}
\def\be{\begin{equation}}
\def\ee{\end{equation}}
\begin{document}

                \title{Integrable System Constructed out of
        Two Interacting Superconformal Fields}

                        \author{by \\
                        Ziemowit Popowicz\\ \\
        Institute of Theoretical Physics, University of Wroc{\l}aw\\
        Pl. M. Borna 9 50 - 205 Wroc{\l}aw Poland}
\date{}
\maketitle

\begin{abstract} We describe how it is possible to introduce  the 
interaction between  superconformal fields of the same conformal 
dimensions. In the classical case such construction can be used to 
the 
construction of the Hirota - Satsuma equation. We construct 
supersymmetric 
Poisson tensor for such fields, which generates a new class of 
Hamiltonin 
systems. We found Lax representation for one of equation in this 
class by supersymmetrization Lax operator responsible for Hirota - 
Satsuma equation. Interestingly  our supersymmetric equation is not 
reducible to classical Hirota - Satsuma equation. We show that our 
generalized system is reduced to the one of the supersymmetric KDV 
equation (a=4) but in this limit integrals of motion are not reduced 
to 
integrals of motion of the supersymmetric KdV equation.
\end{abstract}

\newpage

\section{Introduction. }
The Korteweg - de Vries equation is probably  most popular soliton 
equation which have been extensively studied by mathematicians as 
well as 
by physicists [1] in the last 30 years. Beside supplying nice example 
of the completely integrability of this equation, it bears a deep 
relation 
to conformal field theory [2], 2D gravity and matrix models [3].

A remarkable feature of the KdV hierarchies is its relation, via the 
second 
Hamiltonina structur to the Virasoro algebra discovered by Gervais 
[4] .
This observation has been extended to other Lie algebras also,  
as for example the Nonlinear Schrodinger equation is connected 
with the $SL(2,C)$ Kac -Moody algebra [5] and the Boussinesq equation 
connected with the so called $W_2$ algebra [6].

On the other side many different generalizations of the soliton 
equation 
have been proposed recently as the Kadomtsev-Petviashvilli, 
Gelfand - Diki hierarchies and supersymmetrization. The motivations 
for studing these are diverse, for example in supersymmetric 
generalization, it is the observation, that in the so called bosonic 
limit 
of supersymmetry $(susy)$, sometimes we obtain a new class of the
integrable models. Up to now, supersymmetric KdV hierarchies [7-16] 
have 
been constructed for $ N=1,2,3$ and $4$ based on relation to 
superconformal algebras. For extended $ N=2 $ supersymmetric case the 
Boussinesq [17-18], Nonlinear Schrodinger equation [5,19-21] and 
multicomponent Kodomtsev - Petviashvilli hierarchy [22] have been 
supersymmetrized as well.

It appeared that in order to get a supersymmetric $susy$ theory we 
have to
add to a system of $k$ bosonic equations  $kN$ fermions and $k(N-1)$ 
boson 
fields $(k=1,2,...,N=1,2,...)$ in such a way that final theory 
becomes $susy$
invariant. Interestingly enough, it appeared that during the 
supersymmetrizations, some typical $susy$ effects (compare to the 
classical
theory) occured. We mention a few of them :
the nonuniqueness of the roots for $susy$ Lax operator [15],
the lack of the bosonic reduction to classical equations (for example 
in $susy$ Boussinesq equation [17]) and the occurence of non-local
conservation laws [23]. These effects rely strongly on the 
descriptions of 
the generalized systems of equations which we would like to 
investigate.

In this letter we would like to study problem how it is possible to 
build the Hamiltonian operator (Poisson tensor) and integrable system 
using 
two interacting between themselves (super)conformal fields. That it 
is 
possible to carry out such construction (susy) Boussinesq equation is 
good example, where we have two conformal fields with different 
conformal 
dimensions. However we are interseting in construction where we use 
two different fields with the same conformal dimensions.

First we study classical aspect of our problem without any reference 
to 
supersymmetry and next we consider its supersymmetrizations.

In the "classical" part we show that it is possible to construct 
several 
different Poisson tensors using two conformal fiels of same 
dimensions.
We carried out this construction assuming that in limiting case when 
second field vanishes our Poisson tensor reduces to  Poisson tensor 
which is connected with Virasoro algebra and hence reproduces 
Korteweg-de Vries equation. Among these different Poisson tensors 
there is 
tensor which could be used to construction of Hirota-Satsuma  
equation [24]. This equation is a nontrivial extension of  Korteweg -
de 
Vries equation which is integrable, possesses Lax operator [25] and 
has the recursion operator [26]. Moreover in limiting case of pure 
KdV 
equation (when second field vanishes) integrability is preserved and 
Lax 
operator is reduced to KdV counterparts.

In supersymmetric case, presented in second part, we have much more 
complicated situation compare to classical one. First we carried out
classifications of all possible supersymmetric Poison tensor 
constructed 
out of two superconformal fields of the same dimensions. We used in 
this 
aim the symbolic computer language Reduce [27] and computer package 
SUSY2
[28]. Similarily to the classical case we 
assumed that these tensors should be reducible to tensors connected 
with 
the $N=2$ super Virasoro algebra and hence to those which reproduces 
$susy$ generalizations of $KdV$ equation. The $susy (N=2)$
extension of  $KdV$ equation is a class of equations containing  
one free parameter, however only three  members of this class $ 
a=1,4,-2 $ 
are completely integrable and possesess Lax pairs. Therefore using 
once more computer and package SUSY2 we investigate Lax operator. We
assumed the most general ansats on Lax operator,
constructed out of two superconformal fields, assuming that it 
reduces 
to known Lax operators of $susy KdV$ equations. We showed that 
it reproduces consistent equation only for system which is reduced to 
$susy KdV (a=4) $ equation. Interestingly our Lax operator for that 
system reduces to very simple form. Finally we present three 
nontrivial 
Hamiltoninans for our system.

As the result, in the bosonic limit of our system, we obtained a 
complicated system of four interacting classical fields.
Surprisingly these equations are not reduced to classical 
Hirota - Satsuma equation. We are not surprised, because as we 
mentioned 
earlier for super extension of  Boussinesq equation we encounter the 
same situation - the lack of the proper bosonic limit. There is also 
second 
aspect of our supersymmetrization: namely conservations laws of our 
super system does not coincide, in the limit of pure $susy KdV$ 
equation,
with the conservations laws of $susy$ $KdV$ $(a=4)$ equation.  We 
proved
it by showing the absence of the integrals of motions of second, 
fourth 
and sixth conformal dimension in our generalization. 
Let us remark that our generalization  which
could be considered as the supersymmetrization of the Hirota-Satsuma 
equation is integrable due to the existence of Lax operator.

\section{ Classical Poison tensor and Hirota - Satsuma equation}

Let us start our consideration noticing that famous Korteweg - de 
Vries 
equation 
\be
 u_{t} =  - u_{xxx} + 6uu_{x},
\ee
can be treated as a Hamiltonin system 
\be
u_{t} = \{ u, H \}, 
\ee
with the Hamiltoninan and the Poisson brackets defined by 
\be
H=\frac{1}{2} \int u^2 dx,
\ee
\be
\{u(x),u(y)\}=(-\partial^3 + 2u\partial +2\partial u)\delta(x-y).
\ee
For later pourpose let us rewrite this equation in the equivalent 
form 
using the Poisson tensor
\be
P_{2} = -\partial^3 + 2u\partial + 2\partial u,
\ee
\be
u_{t} = P_{2}\,\, grad(H),
\ee
where $grad$ denotes the functional gradient. 

For the Fourier modes of $u(x)$,
\be
u(x)=\frac {6}{c}\sum_{n=-\infty}^{\infty}exp(-inx)L_{n} -\frac 
{1}{4},
\ee
the Poisson brackets in eq. (4) imply the structure relations of the
Virasoro algebra 
\be
[L_n,L_m] = (n-m)L_{n+m}  + cn(n^2-1)\delta_{n,m}, 
\ee
where c is a central extension term.

It is well known that this 
equation is completely integrable with infinite numbers of 
integrals of motion being in invotution among themselves.
The interesting problem in theory of solitons is to generalize the KdV
equation to system of equations, in such a way, that to preserve 
integrability and in limiting case, where additional fields vanishes, 
to 
recover the usual Korteweg de Vries equation. 
At the moment we have many different proposal and one of them 
is the utilization of Poisson tensor constructed out of two different 
conformal fields $u$ and $w$ of same conformal dimensions.
Taking into account that the field $u$ is two dimensional, while 
usual 
Poisson tensor of KdV equation is three dimensional, we make the 
following 
ansatz \be
P_2=\pmatrix{ c_1\partial^3_{x} + z_1kd(u) &
               c_2\partial^3_{x} + z_2kd(u) + z_3kd(w) \cr
          c_2\partial^3_{x} + z_2kd(u) + z_3kd(w) &
        c_3\partial^3_{x} +z_4kd(u) + z_5kd(w)} , 
\ee
where $c_1,.. z_1,...$ are at the moment free coefficients and 
\be
kd(u)=u\partial_{x} + \partial_{x} u.
\ee
In order to obtain the conditions on the coefficients $c_{i}$ and 
$z_{i}$ 
we verified the Jacobi identity [29]
\be
<a,P^{'}[Pb]c> + cyclic{~}permutation{~}of{~}(a,b,c) =0 ,
\ee
where $P^{'}[Pb]$ denotes the directional derivative along $Pb$ and 
$<,>$ 
is a scalar product while $a,b,c$ are arbitrary test functions.
We obtained three different solutions on the coefficients $c_{i}$ and 
$z_{i}$
\be
    z_{2}=z_{3}=z_{4}=c_{2}=0,
\ee
\be
    z_{3}=0, {~} \
    z_{5}=\frac {z_{2}^2 - z_{1}z_{4}}{z_{2}},{~} \
  c_{1}=\frac {c_{2}z_{1}}{z_2},
\ee
\be
 z_{2} = 0 ,{~} \
 z_{1} = z_{3},{~} \
 c_{3} = \frac {c_{1}z_{4}+c_{2}z_{5}}{z_{3}},
\ee
The first solution give us the direct product of two standard 
Virasoro structures eq.(5) with arbitrary central charges $c_{1}$ 
and $c_{3}$. We can apply this Poisson  tensor to gradient of 
\be
H=\frac {1} {2} \int uw dx,
\ee
and obtain equation which have been considered in [] in context of 
the extended supersymmetric $(N=3)$ KdV system. We will not consider 
further such possibilities and concentrate our attention on other 
solutions.

The second solutions is not interesting from our point of view 
because it
impossible to reduced, in the usual manner, this tensor to the 
standard 
Virasoro type Poisson tensor. Indeed in order to see it let us brifly 
explain standard Dirac reduction [6] formula. 

Let $U,V$ be two linear spaces with coordinates $u$ and $v$. Let 

\be 
P(u,v)=\pmatrix{ P_{uu}, & P_{uv} \cr
        P_{vu} , & P_{vv}},
\ee
ba a Poisson tensor on $ U \bigoplus V $ . Assume that $P_{vv}$ is 
invertible, then 
\be
P=P_{uu} - P_{uv}P_{vv}^{-1}P_{vu},
\ee
is a Poisson tensor on $U$.

As we see the reduction, for  second solution, in space where $w=0$ 
is possible if $c_{2}=0$ and $z_{2}=0$, but then we obtain undefined 
central 
extension term. However it is interesting to notice that 
we can carry out the reduction in  different manner also. Indeed we 
can 
deform this structure, in the following self-consistent way:
\be
w \rightarrow z_{2}w,{~}{~}{~}
\frac {c_{2}} {z_{2}} \rightarrow k,{~}{~}{~} c_{2} \rightarrow 0,{~} 
{~}{~}
 z_{2} \rightarrow 0,
\ee
and obtain desired result.
On the other side it is possible to make reduction in  space where 
$ u=0 $ assuming that $c_{2} =0$ and obtaining  standard Virasoro 
type 
tensor for the field $w$. In the next we will not consider this case 
also.

The last solution is most interesting which allows us to make  
reduction in the space $w=0$ assuming that $ c_{2}=0$. 
This class of Poisson tensor includes the Hamiltonian operator 
responsible 
for the Hirota - Satsuma equation which has the form
\be
P_{2}=\pmatrix{ \partial^3 +\partial u + u\partial & 
\partial w + w\partial \cr
 \partial w + w \partial 
 & 2\partial^3 +2\partial u +2u\partial},
\ee
From this form of  Poisson tensor we see that the interaction of the 
fields is concentrated in the the diagonal as well 
as off diagonal elements of tensor. Therefore we can state, that 
we constructed extended Virasoro algebra which contains usual 
conformal algebra interacting with additional conformal field. 

The Hamiltonian and equations of motion for Hirota -Satsuma system are
\be
H=\frac{1}{2} \int u^2 - w^2 ,
\ee
\begin{eqnarray}
u_{t} &=&  u_{xxx} + 3uu_{x} - 3ww_{x}\, ,\\
w_{t} &=& -2w_{xxx} - 3uw_{x}\, .
\end{eqnarray}
Hirota and Satsuma have found [24]  five notrivial 
integrals of motion and latter it was proved that this equation 
is integrable by presenting its Lax representations [25]
\be
L= (\partial^2 + u + w)*(\partial^2+u-w),
\ee
\be
L_{t}= [ L,(L^{\frac {3}{4}})_{+}],
\ee
where $(+)$ denotes the projection onto the pure differentail part 
of the operator.

\section{The extended supersymmetrization of Poisson tensor 
constructed 
out of two fields. }

        The basic objects in the supersymmetric analysis are the 
superfield
and the supersymmetric derivative. We will deal with the so called 
extended 
$N=2$ supersymmetry for which superfields are superfermions 
or superbosons depending,  in addition to 
x  and t, 
    upon two anticommuting variables,  
    $\theta_{1}$     and     
      $\theta_{2}$,  
      $(  \theta_{2}\theta_{1} = -   \theta_{1}
\theta_{2},   \theta_{1}^{2} =   \theta_{2}^{2} = 0)$.
Their Taylor expansion with respect to $\theta$ is
\be
U(x,   \theta_{1},  \theta_{2})
= u_{o}(x) + \theta_{1} \zeta_{1}(x) + 
  \theta_{2}\zeta_{2}(x) + 
    \theta_{2}\theta_{1}u_{1}(x),
    \ee
    where the fields   
    $u_{o}, u_{1}$,    are to be interpreted as the boson (fermion) 
fields
for superboson (superfermion) field, while     
$\zeta_{1}, \zeta_{2}$,  as fermions (bosons) 
for superboson (superfermion) respectively. The superderivatives are 
defined as
\be
{\cal D}_{1} =\partial_{\theta_{1}} + \theta_{1}\partial \, ,
\qquad \qquad
{\cal D}_{2} =\partial_{\theta_{2}} + \theta_{2}\partial,
\ee
with the properties
\be
{\cal D}_{2}{\cal D}_{1} +
{\cal D}_{1} {\cal D}_{2} = 0 \, ,
\qquad \qquad
{\cal D}_{1}^{2} = {\cal D}_{2}^{2} = \partial.
\ee
        Below we shall use the following notation: $({\cal
D}_{i}{F})$ denotes the outcome of the action of superderivative on 
the 
superfield, while ${\cal D}_{i} {F}$ denotes  action itself.

The supersymmetric Poisson tensor connected with 
Virasora algebra has the form
\be
P=cD_{1}D_{2}\partial + zs(U),
\ee
\be
s(U)=2\partial U + 2u\partial - D_{1}UD_{1} - D_{2}UD_{2},
\ee
where $c$ is central extension term, while $z$ an arbitrary free 
parameter.
We assume that in $susy$ case the analog of the formula (9)
reads
\be
P_{2}=\pmatrix{c_{1}D_{1}D_{2}\partial + z_{1}s(U)& 
c_{2}D_{1}D_{2}\partial + z_{2}s(U)+  z_{3}s(W) \cr
c_{2}D_{1}D_{2}\partial + z_{2}s(U) + z_{3}s(W)&
c_{3}D_{1}D_{2}\partial + z_{4}s(U) + z_{5}s(W)}.
\ee
We checked the Jacobi identity using the same formula as in classical 
case and got the same conditions on the central extensions terms 
$c_{i}$ and 
$z_{i}$ as in classical case. From same reasons, as in classical 
case, we consider last solution only, assuming additionally that 
$c_{2}=0$. 
We can easily obtain some Hamiltonian system, using analog of formula 
(6) 
in which we consider the most general three conformal dimensional 
Hamiltonian.
Such Hamiltonian has following density
\be
\begin{array}{ll}
H= & a_{1}(D_{1}D_{2}U)U + a_{2}(D_{1}D_{2}U)W + a_{3}(D_{1}D_{2}W)W 
+ 
\cr\cr
&
a_{4}W^3 + a_{5}W^2U + a_{6}WU^2 +a_{7}U^3 + a_{8}W_{x}U/, ,
\end{array}
\ee
where $a_{i}$ are an arbitrary coefficients, superboson  $U$ is 
defined by eq.(25) while superboson $W$ is 
\be
W=w_{o} + \theta_{1}\xi_{1} + \theta_{2}\xi_{2} +
 \theta_{2}\theta_{1}w_{1},
\ee
where $\xi_{i}$ are the fermions valued functions, while $w_{i}$ are 
classical functions.

In that manner it is possible to obtain a hudge class 
of complicated Hamiltonian systems, with many free parameters and this
which contains the $susy$ generalization of the $KdV$ equation.

\section{The strategy and results.}

We have seen in the last section that it is possible to obtain
new class of Hamiltonin systems. We would like to find, in this 
class, 
the integrabe one and this which contains the Hirota - Satsuma 
equation in 
bosonic limit. Therefore we applied following strategy in order to 
solve this problem:

1.) We assume the equations of motion on the superbosons $U$ and $W$ 
in 
the form which is obtained by applications of Poisson tensor (30) 
with the conditions (14) and $c_{2}=0$ to the gradient of Hamiltonian 
(31).

2.) We construct most general $susy$ generalization of "classical"  
Lax 
operator appearing in Hirota-Satsuma equation (23) and investigate  
supresymmetric generalization of its Lax pair (24).

3.) We use equations of motion constructed in first approach to the 
verifaction of validity of Lax pair obtained in second approach. In 
that 
manner we obtain the system of algebraic equations on 
the free parametrs which appeare in  Lax operator as well as in 
equations of motion and in Poisson tensor. This system of equation we 
would 
like to solve.

Before presenting  the results of our computations let us brifly 
recall 
basic facts on  $susy$ $N=2$ generalizations of the KdV equation which
are needed in this construction. This generalization could be written 
down as
\be
\begin{array}{ll}
U_{t} =& P\,grad(\frac{1}{2}U(D_{1}D_{2}U) + \frac{a}{3}U^3) =
\cr\cr
&
\partial(-U_{xx} + (2+a)U(D_{1}D_{2}U) +(a-2)(D_{1}U)(D_{2}U) + aU^3) 
\, ,
\end{array}
\ee
where $P$ is defined by (28) while $a$ is an arbitrary parameter. It 
appeared that this $susy$ generalisation is integrable only for three 
values
of parametres $a$. The integrability have been  
conclude from the observation that it is possible to 
find Lax operators [10,16] for these cases.

The Lax operator in the supersymmetric case is an element of the 
super 
pseudo-differential algebra $G$ which each element $g$ could be 
presented as
\be
G \ni g =\sum_{n=-\infty}^{\infty}\Phi_{n}\partial^{n}=
 \sum_{n=-\infty}^{\infty}(B_{n} + F_{n}D_{1} + FF_{n}D_{2} +
    BB_{n}D_{1}D_{2})\partial^n,
\ee 
where $B_{i}$ and $BB_{i}$ are arbitrary superbosons while $F_{i}$ 
and 
$FF_{i}$ are arbitrary superfermions. In our case of  $susy$ KdV
generalization, Lax operators are given by:

\begin{eqnarray}
a=-2 : L &=&\partial^2 + D_{1}UD_{2} - D_{2}UD_{1} \, , \\
a=4\,\,\,\,  : L &=&\partial^2 -(D_{1}D_{2}u) - u^2 + (D_{2}u)D_{1} - 
        (D_{1}U)D_{2} - 2UD_{1}D_{2} , \nonumber\\
         &=&-(D_{1}D_{2} + U)^2 \, , \\
a=1\,\,\,\, : L &=&   \partial - \partial^{-1}D_{1}D_{2}U \, .
\end{eqnarray}

For first two cases we have usual Lax pair [10]
\be
    \frac{\partial U}{\partial t} = 4 [ L, L^{\frac{3}{2}}_{+}] \, ,
\ee
while for the last case we have nonstandard Lax pair [16]
\be 
    \frac{\partial u}{\partial t} = [L, L^{3}_{\leq 1} ] \, ,
\ee
where $L^{3}_{\leq 1}$ denotes the projection on the subspace of the 
\be
P_{\leq 1}(\Gamma) =\sum_{n=1}^{\infty}\Phi_{n}\partial^{n} +
(F_{0}D_{1} + FF_{0}D_{2} +BB_{0}D_{1}D_{2}) \, .
\ee

We do not consider, in the next, nonstandard representation, because 
we
do not have such for the Hirota-Satsuma equation.

We have odd and even dimensional integrals of motion for $a=4$ case. 
Odd 
integrals contains usual conservations of law of $KdV$ equation while 
even 
integrals does not have such property. Explicitely we have first four 
integrals of motion for $a=4$ case
\begin{eqnarray}
I_{1} &=&\int U dxd\theta_{1}d\theta_{2}\, ,\\
I_{2} &=&\int U^2 dxd\theta_{1}d\theta_{2}\, ,\\
I_{3} &=&\int ((D_{1}D_{2}U)U +\frac{4}{3} U^3) 
dxd\theta_{1}d\theta_{2}\, 
\\
I_{4} &=&\int (U_{x}^2 + 3(D_{1}D_{2}U)U^2 
+2U^4)dxd\theta_{1}d\theta_{2}\, .
\end{eqnarray}
These integrals could be computed using following formula
\begin{eqnarray}
I_{2k+1} &=&\int Tr L_{1}^{2k+1} dxd\theta_{1}d\theta_{2}\, ,\\
I_{2k}  &=&\int Tr (L_{1}L_{2})^k dxd\theta_{1}d\theta_{2}\, ,
\end{eqnarray}
where $Tr$ denotes trace formula defined on the 
algebra of $susy$ pseudo-di\-ffe\-re\-ntial algebra $G$. We use usual 
definition of $Tr$ as this which denotes element standing before 
$D_{1}D_{2}\partial^{-1}$ in the algebra $G$. $L_{1}$ and $L_{2}$ are 
two 
different roots of Lax operator (eq. 33) where $L_{1}$ has standard 
form as 
$\partial +...$, while $L_{2}$ is $D_{1}D_{2} + U$.

In order to construct  Lax operator for our generalization we assumed 
that it has following reprsentation
\be
L=\partial^4 + \Phi_{3}\partial^3 + \Phi_{2}\partial^2 + 
\Phi_{3}\partial + 
\Phi_{4} \, ,
\ee
where $\Phi_{i}$ are $susy$ operators of i-th conformal
dimension constructed out of all possible 
combinations of $D_{1},D_{2},D_{1}D_{2},u,w$, $(susy)$de\-ri\-va\-ti\-
ves of 
$u,w$ and with free parameters. It is a hudge expression which 
contains 
243 terms (or in other words 243 free parametrs). We had made two 
additional 
assumptions:

First: in the limit $W=0$ we should recover Lax operator for $susy$
KdV equation in the form of eq.35 or eq.36

Second : our ansatz should be $O(2)$ invariant under the change of 
the 
supersymmetric derivatives $(D_{1} \mapsto -D_{2}, D_{2} \mapsto 
D_{1})$. 
This invariance follows from physical assumption on the nonprivliging 
the 
"fermionic" coordinates in the superspace.

These assumptions simplify our ansats on Lax operator giving for  
$a=4$ 
case 208 terms while for $a=-2$ case 195 terms only. After making 
these 
simplifications we were able to realized third point in our strategy 
and it appeared that only for $a=4$ case we  solved our consistency 
conditions and obtained one nontrivial solution only. Our system of 
equation 
could be written down as
\be
\frac{d}{dt}\pmatrix{U \cr\cr W} =P_{2}*grad((D_{1}D_{2}U)U+
\frac{4}{3}U^3+(D_{1}D_{2}W)W - 2W^2U),
\ee
where 
\be
P_{2}=\pmatrix{ D_{1}D_{2}\partial + s(U) &
s(W) \cr\cr s(W) & D_{1}D_{2}\partial + s(U)},
\ee
and $s(U)$, $s(W)$ are defined by (29).

Explicitely we obtained
\begin{eqnarray}
U_{t} &=& \begin{array}[t]{l}\partial [- U_{xx} + 3(D_{1}U)(D_{2}U) 
+6(D_{1}D_{2}U)U + 4U^3 + \cr
3(D_{2}W)(D_{1}W) - 6W^2U ]\, ,
\end{array}\\
W_{t} &=& \begin{array}[t]{l}\partial [-  W_{xx} - 2W^3 + 
3(D_{2}W)(D_{1}U) 
- 3(D_{1}W)(D_{2}U)]- \cr
 6(D_{2}W)(D_{2}U)U - 6(D_{1}W)(D_{1}U)U \, .
\end{array}
\end{eqnarray}
In the bosonic sector we obtained
\begin{eqnarray}
u_{ot}&=&\partial [ - u_{oxx} + 6u_{1}u_{o} + 4u_{o}^{3}-6w_{o}u_{o} 
],\\
w_{ot}&=&\partial [ - w_{oxx} -2w_{o}^{3} ],\\
u_{1t} &=& \begin{array}[t]{l}\partial [-   u_{1xx} + 3u_{1}^2 + 
3w_{1}^2 + 
           3w_{ox}^2 - 3u_{ox}^2 - 6u_{oxx}u_{o} + \cr
          12u_{1}u_{o}^2- 6w_{o}^2u_{1}  - 12w_{1}w_{o}u_{o}] \, ,
           \end{array} \\
w_{1t} &=& \begin{array}[t]{ll} \partial [ -  w_{1xx} + 6w_{ox}u_{ox} 
+ 
            6w_{1}u_{1} - 6w_{1}w_{o}^2 ] \cr
            +12w_{1}u_{o}u_{ox} - 12 w_{ox}u_{1}u_{o} \, .
            \end{array}
\end{eqnarray}
Interestingly our Lax operator has simple representation
\be
L:=[(D_{1}D_{2} + U + W)(D_{1}D_{2} + U -W)]^2 \, ,
\ee
This form of Lax operator suggests to consider much simpler Lax pair, 
namelly
it is enought to investigate the root of this Lax operator
\be
L=(D_{1}D_{2} + U + W)(D_{1}D_{2}  + U - W) ,
\ee
with corresponding Lax pair
\be
\frac{d\, L}{d\, t} =-4i[L,(L^{\frac{3}{2}})_{+}].
\ee

If we further reduce bosonic limit of our system of equation, 
demanding that 
$u_{o}=0$ and $w_{o}=0$ we obtain the following system
\begin{eqnarray}
u_{1t} &=& \partial(- u_{1xx} + 3u_{1}^2 + 3w_{1}^2)\, ,\\
w_{1t} &=& \partial(- w_{1xx} + 6u_{1}w_{1})\, ,
\end{eqnarray}
which does not coincides with the Hirota-Satsuma equation  eq. (21 - 
22).
Moreover we can transform last equations to the system of 
noninteracting 
two Korteweg - de Vries equations using
\begin{eqnarray}
u_{1} &\mapsto & u_{1} + w_{1} \, , \\
w_{1} &\mapsto & u_{1} - w_{1} \, . 
\end{eqnarray}
However we can not state the same on the supersymmetric level.

It is rather unexpected result because our supersymmetrization method 
used 
supersymmetrizations of the Hirota - Satsuma Lax operator. The 
observation 
that in the process of the supersymmeytrization we do not obtain, in 
bosonic 
limit, the desired equation, is known in the theory of 
supersymmetrization 
of soliton's equation. It happen for example in $susy$ Boussinesq 
equation. 

Finally let us discuss the problem of existence of the integrals of 
motion 
in our model. We successed to construct first three conservations of 
laws 
which are

\begin{eqnarray}
I_{1} &=&\int U dxd\theta_{1}d\theta_{2}\, , \\
I_{3} &=&\int ((D_{1}D_{2}U)U + \frac{4}{3}U^3 + (D_{1}D_{2}W)W -
2W^2U) dxd\theta_{1}d\theta_{2}\, , \\
I_{5} &=& \begin{array}[t]{l} \int ( 16U^5 +40(D_{1}D_{2}U)U^3 
+10(D_{1}D_{2}U)^2U +30U_{x}^2U - 
\cr\cr
5(D_{1}D_{2}U_{xx})U -10(D_{1}D_{2}W)W^3 - 5(D_{1}D_{2}W_{xx})W +
\cr\cr
20(D_{2}W_{x})(D_{2}W)U +
20(D_{1}W_{x})(D_{1}W)U + 50(D_{2}W)(D_{1}W)U^2+ 
\cr\cr
20W_{xx}WU +
 30(D_{1}D_{2}W)^2U - 30(D_{1}D_{2}W)WU^2 + 
20W_{x}^2U+
\cr\cr
30W^4U 
- 3(D_{1}D_{2}U)W^2U-40W^2U^3) dxd\theta_{1}d\theta_{2},
\end{array}\, ,
\end{eqnarray}

Moreover we proved the absence of integrals of motion of 
second,fourth and 
sixth conformal dimensional in our system. It is also unexpected 
result. It 
means that our equations of motion does not coincide, in the limit 
when 
$W=0$ with $susy$ version of $KdV$ equation, because as we saw the 
last one 
possesses odd and even conformal dimensional integrals of motions. 
In order to explain this situation let us make two observation:

First: We introduced interaction by including second fields and this 
field 
destroys "half" of integrals of motion for $susy$ $kdV$ equation.

Second: For pure $susy KdV a=4$ case  Lax operator possesses  two 
nonequivalent roots. These roots are responsible for the integrals 
of motion. In our case we have no such situation and we have one root 
of 
Lax operator only.

\end{document}